\documentclass[12pt]{article}
\usepackage{geometry}
\usepackage{natbib}
\usepackage{framed}

\usepackage{amsgen,amsmath,amstext,amsbsy,amsopn,amssymb,amscd}
\usepackage{graphicx}
\usepackage{adjustbox}
\usepackage{caption}
\usepackage{subcaption}
\usepackage{float}
\usepackage{algorithm}
\usepackage{algpseudocode}
\usepackage{setspace}
\usepackage{mathtools}
\usepackage{enumitem}
\usepackage{multirow}
\usepackage{verbatim}

\usepackage{adjustbox}
\usepackage{amsthm}
\usepackage{appendix}

\newtheorem{theorem}{Theorem}
\newtheorem{lemma}{Lemma}[theorem]
\newtheorem{corollary}{Corollary}[theorem]

\newtheorem*{remark}{Remark}
\newtheorem{definition}{Definition}[section]

\newcommand{\one} {{\boldsymbol{1}}}

\newcommand{\dsum}{\displaystyle\sum}

\newcommand{\tab}{\hspace{0.5cm}}

\newcommand{\given}{\, | \,}
\newcommand{\st}{\, : \,}

\newcommand{\RR}{{\bf R}}
\newcommand{\PP}{{\bf P}}
\newcommand{\XX}{{\bf X}}
\newcommand{\FF}{{\bf F}}

\DeclareMathOperator{\Cor}{cor}
\newcommand{\corh}[1]{\widehat{\Cor}_n\left(#1\right)}
\newcommand{\VV}{{\bf V}}
\newcommand{\DD}{{\bf D}}
\newcommand{\SSS}{{\bf S}}

\newcommand{\bas}{\begin{align*}}
\newcommand{\eas}{\end{align*}}
\newcommand{\WW}{{\bf W}}
\newcommand{\YY}{{\bf Y}}
\newcommand{\UU}{{\bf U}}
\newcommand{\reals}{\mathbb{R}}

\DeclareMathOperator{\Cov}{Cov}
\DeclareMathOperator{\Var}{Var}
\newcommand{\cov}[1]{\Cov \left(#1\right)}

\newcommand{\var}[1]{\Var \left(#1\right)}

\newcommand{\dia}{\hat{\Delta}(i,A)}
\newcommand{\p}{\mathrm{p}}

\begin{document}


\begin{title}
{\Large\bf A testing-based approach to the discovery of differentially correlated variable sets.\\
}
\end{title}
\author{Kelly Bodwin, Kai Zhang, and Andrew Nobel}
\date{\today}
\maketitle


\begin{abstract}
Given data obtained under two sampling conditions, it is often of interest to identify variables that behave differently in one condition than in the other. 
We introduce a method for differential analysis of second-order behavior called Differential Correlation Mining (DCM). The DCM method identifies differentially correlated sets of variables, with the property that the average pairwise correlation between variables in a set is higher under one sample condition than the other. 
DCM is based on an iterative search procedure that adaptively updates the size and elements 
of a candidate variable set. Updates are performed via hypothesis testing of individual variables, 
based on the asymptotic distribution of their average differential correlation. 
We investigate the performance of DCM by applying it to simulated data as well as recent experimental datasets in genomics and brain imaging.

\end{abstract}
%
%
%
%
\vfill

\small{Address for correspondence:  Kelly Bodwin, \emph{kbodwin@email.unc.edu}}


\pagebreak

\section{Introduction}
\label{intro}

In many statistical problems one has two datasets that measure the same variables under different conditions. In the analysis of such data, it is common to assume that the samples in each dataset are generated from two underlying distributions.  Even when the data is high dimensional, differences between the distributions may only be present for a small number of variables, and it is often of interest to identify these key variables. In this paper, we present a new method of second order comparative analysis, called Differential Correlation Mining (DCM), which identifies sets of variables such that the average pairwise correlation between variables in the set is higher in one sample condition than in another. The method does not make use of auxiliary information, apart from the separation of samples into pre-determined groups (e.g. treatment vs control). DCM is applicable to both low and high dimensional datasets. 

Most often, differential behavior between sample groups is measured by \emph{first-order} statistics, which are functions of a single variable. Familiar first-order statistics include the sample mean and the sample variance. A well-studied example of first-order differential analysis is the study of differential gene expression in microarrays (see \citet{de1} for a canonical example, or \citet{de2} and the references therein for an overview of several methods). Other applications of first-order differential analysis include text analysis for authorship identification \citep{authorship}, studies of brain functionality based on regional activation \citep{diffbrains}, and investigation of cultural bias in standardized testing \citep{testbias}. 

The use of first-order statistics only allows for analysis of a single variable at a time. To study relationships between pairs of variables, one requires functions of two variables, which specify \emph{second-order} statistics. Examples of second-order statistics include correlation, covariance, and distance.  When one wishes to understand interactions between many variables (as in clustering problems), data may be summarized in matrix form, where each entry in the matrix represents the observed value of a second-order statistic. It is common to look within a matrix of relational data for groups of variables that have high pairwise association.  Interconnected variable sets may represent, e.g., social groups in a communication network \citep{facebook}, genes in common protein pathways \citep{geneclust}, functionally similar brain regions \citep{brainclust}. 

While there is a large literature on clustering and networks, to the best of our knowledge, there is relatively little work comparing second-order behavior across two sample conditions. The many insights obtained from ordinary second-order variable set selection lead us to believe that a second-order differential approach will be of scientific interest. The methods introduced in this paper fall under the broader heading of \emph{differential association mining}.  As in ordinary association mining, we are interested in the pairwise behavior of variables; however, in the differential setting, we must consider two different relational matrices. In some cases, simply taking the difference of the matrices and applying ordinary clustering methods would suffice. However, most second-order statistics - including the focus of this paper, the linear correlation coefficient - require a more careful treatment.  For instance, two sample correlation matrices will exhibit vastly different random behavior based on the sample sizes of the corresponding datasets, and will have a complex dependency structure when the population correlation matrix is not the identity. 

The DCM method proposed here addresses differential correlation mining in a direct way. (Section \ref{exist} considers possible alternatives based on existing work.) DCM seeks variable sets that form differentially correlated (DC) cliques. In a graph, a clique is a set of nodes that is fully connected, in the sense that there is an edge between every pair of nodes in the set. Informally, a DC clique is a set of variables such that each variable in the set has a positive (usually large) average differential correlation with the other variables in the set.

More formally, let $\RR_1, \RR_2$ be the $p \times p $ population correlation matrices of the distributions underlying sampling conditions 1 and 2, respectively. 
Let $J \subset [p]$, where $[p]$ is the index set $\{ 1, ..., p \}$, and define
\begin{equation}
\label{eq:delta}
	\Delta(i,J)  \, = \, \frac{1}{|J|}  \dsum_{j \in J} \left( \RR_1- \RR_2 \right)_{ij}
\end{equation}
to be the difference of correlations between variable $i$ and variables in index set $J$. 
Here the subscript $ij$ denotes the element in the $i$-th row and $j$-th column of the corresponding matrix, and $|J|$ is the cardinality of the set $J$.  We define DC cliques as follows.

\begin{definition} \label{def:DCc} 
Let $\RR_1, \RR_2$ be given and let $\Delta(\cdot, \cdot)$ be defined as in (\ref{eq:delta}).
An index set $A \subseteq [p]$ with at least two elements is a \emph{DC clique} for $\RR_1 - \RR_2$ if

\begin{enumerate}
\item $\Delta(i, A) > 0$ iff $i \in A$,
 \item The set \emph{A} cannot be written as a disjoint union of nonempty index sets $A_1, A_2 \subset [p]$ such that $A_1$ and $A_2$ satisfy condition 1 above.
\end{enumerate}

\end{definition}

Condition 1 ensures that no relevant variables are omitted from a DC clique 
(every variable that is positively differentially correlated relative to the set $A$ is included in $A$) and that 
a DC clique does not contain any extraneous elements. 
Condition 1 implies that a DC clique 
has larger average pairwise correlation under the first distribution than under the second. 
Condition 2 ensures that a DC clique cannot be subdivided into two smaller DC cliques.
Importantly, the definition places {\em no conditions} on the correlation matrices $\RR_1$ and $\RR_2$.
In particular, $\RR_1$ and $\RR_2$ need not be sparse, and need not satisfy any structural constraints such as bandedness. 
For a given pair $\RR_1, \RR_2$, it may happen that no DC cliques exist, or that the entire variable set forms a DC clique. Note that the definition of DC cliques is not symmetric: in general, the DC cliques for $\RR_1 - \RR_2$ will be different from those for $\RR_2 - \RR_1$.

As defined above, DC cliques are features of the underlying population distributions of the data. 
The broad objective of DCM is to use observed data to identify DC cliques, or approximations of these, 
without prior knowledge of the identity, number, or size of the DC cliques present in the population. 
It is worth noting that the DCM algorithm and supporting analysis described here is easily adapted 
to a non-differential correlation mining algorithm. Code for the correlation mining procedure is available in the DCM packages for R and Matlab.

\begin{remark} \emph{Some bioinformatics literature uses the phrase ``Differential Co-Expression", sometimes abbreviated ``DC", as an umbrella term for all differential second-order gene expression behavior. In this paper, ``DC" will refer specifically to differential correlation; when a distinction must be made with co-expression or covariance, this will be made explicit.}\end{remark}

\subsection{An example}

\begin{figure}[H]
\includegraphics[width = 0.8\textwidth]{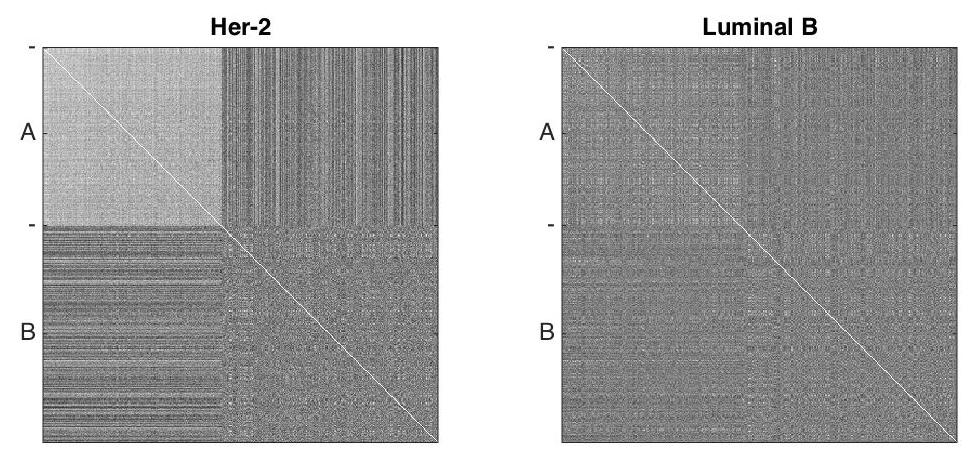}
\centering
\caption{Sample correlation matrices for each of two breast cancer tumor subtypes, showing observed DC clique (A) and random genes (B) .}
\label{fig:example}
\end{figure}

To motivate our definition of DC cliques, we provide an illustrative real-world example. Figure \ref{fig:example} shows an empirical DC clique identified by DCM in real data from The Cancer Genome Atlas (TCGA) Research Network ({http://cancergenome.nih.gov/}). The two sample conditions under consideration are Her-2 type breast cancer tumors and Luminal B type tumors, as classified by \cite{perou}. 
(Further results for the TCGA dataset are provided in Section \ref{realdata}.)

Figure \ref{fig:example} shows the sample correlation matrices within each tumor type, 
restricted to a set of 365 variables consisting of an empirical DC clique of size 165 
selected by DCM ($A$), and 200 randomly chosen variables ($B$). 
The variables $B$ are included for contrast, and to show that the differential correlation 
observed in $A$ is not present in the entire dataset. The figure illustrates the second-order behavior and the differential nature of the identified DC clique $A$. 
The block pattern in the upper left corner of the Her-2 matrix shows that every entry in the correlation 
matrix of $A$ is large, suggesting that all the variables of $A$ are strongly pairwise correlated. 
The Luminal B sample correlation shows a similar pattern, but it is much less pronounced. No such pattern is seen among 
the variables in $B$.

\begin{figure}[H]
\includegraphics[width = \textwidth]{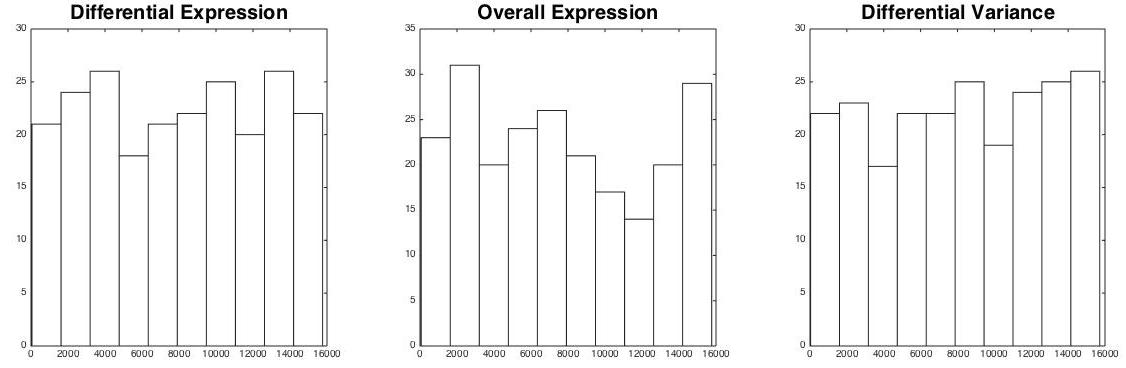}
\caption{Ranks of genes in observed DC clique (A) out of 15,785 total genes. 
\\ \footnotesize{(Ranked by: Differential expression, as measured by p-values of 2-sample t-tests; mean overall expression among Her-2 samples; and difference of sample variance between Her-2 and Luminal B.)}}
\label{fig:diagnostic}
\end{figure}

In general, the results of DCM are distinct from those found by first-order analysis (e.g. differential expression). 
For example, Figure \ref{fig:diagnostic} shows the relative differential expression, overall expression level, and differential variation for the above estimated DC clique $A$. For this plot, we ranked all genes in the study ($p = 15,785$) by (a) $t$ statistic of differential mean expression between Her-2 and Luminal B samples, (b) overall expression in Her-2 samples, and (c) ratio of sample variations ($F$ statistic) for Her-2 versus Luminal B samples. 
The histograms in Figure \ref{fig:diagnostic} show the ranking of the genes in $A$. 
The overall uniformity of the histograms indicates that the variables in the observed DC clique $A$
do {\em not} exhibit standard first-order differential behavior.
Similar results were observed for all other data studied in this paper.

By targeting DC cliques, the DCM method identifies variables whose \emph{joint} behavior is different across sample conditions. The results are readily interpretable as sets of variables that interact strongly under one sample condition but only weakly (or not at all) under another. In this paper, we will demonstrate DCM is an effective and efficient way to identify differentially correlated variable sets from observed data.

\subsection{Related work}

\label{exist}

There is a substantial body of recent work concerning estimation and testing of covariance and 
correlation matrices in high-dimensional settings. 
However, to the best of our knowledge, none of this work addresses the search for DC cliques
or related differential structures in a direct way.  Below we provide an overview of work that
is related to DCM.
In what follows, let $\RR_1, \RR_2$ denote the population correlation matrices of two data 
distributions, and let
$\widehat{\RR}_1, \widehat{\RR}_2$ denote the corresponding sample correlation matrices.

\emph{Mining from single correlation matrices.} Non-differential correlation mining, in which one 
searches for highly associated variables from a single dataset, has been well-studied, typically
in the context of clustering.
\citet{corrclust} provides a survey of clustering methods for high-dimensional data based on 
correlation distance. 
\citet{survey} and \citet{geneclust} and the references therein give an overview of
methods developed specifically for clustering of gene expression. 
In general, typical clustering or community detection methods must be adapted 
for application to correlation distances to correct for bias (see e.g. \citet{commdet} for an illustrative example).

\emph{Estimation and hypothesis testing.}   
There has been a great deal theoretical work devoted to testing equality of 
high-dimensional covariance and correlation matrices. 
When the sample size $n$ is substantially larger than the dimension $p$, classical results are applicable, 
e.g., likelihood ratio tests as discussed in \citet{anderson} and \citet{muirhead}, or results like those 
of \citep{steiger} for testing individual sample correlation. In the high-dimensional ($p > n$) setting, 
\citet{Cai_Covmat, Cai_limiting, Cai_Yin} have developed minimax rate optimal tests for the equality 
of covariance matrices under sparsity assumptions. 
Results for correlation (rather than covariance) are less prevalent; recent work includes tests for sets of 
sample correlation coefficients \citep{equality}, tests for rank-based correlation matrices \citep{rankbased}, 
and tests for detecting overall dependence \citep{hero}.

In some cases, optimal testing procedures can inform methods for estimation of high-dimensional 
covariance and correlation matrices. Particularly relevant is the work of \citet{caizhang}, 
which yields an estimator for the difference matrix $\DD = \RR_1 - \RR_2$. This estimator 
is implemented and discussed further in Section \ref{sims}. 
Other approaches to high-dimensional estimation include the work of \citet{bickel}, who discuss a thresholding estimator for covariance matrices; \citet{partial}, who estimate partial correlations in sparse regression models; 
and \citet{covest}, who make use of graphical model techniques for covariance matrix estimation. 

\emph{Detection of isolated changes in correlation structure.}  
Existing approaches to differential correlation mining are based on examining individual variables 
for changes in second-order structure across two sample conditions. 
For example, one may treat $\widehat{\RR}_1$ and $\widehat{\RR}_2$ as the adjacency matrices 
of two fully connected, weighted networks, and then look for variables whose connectivity pattern is very different across the two networks \citep{yin2, gill}. 
Most methods approach differential correlation mining by developing a statistic to measure 
the change in pairwise correlations of an individual variable. 
\citet{Nstat} uses the covariance distance (total difference of covariances); \citet{gsca} uses a direct difference of sample correlations; \citet{diffcorr} uses the difference of Fisher transformed sample correlations; and \citet{dcgl} use a filtration (or thresholding) step before summing square correlation differences. These methods then permute samples across the two classes to measure the significance of the original differential correlation. Significant variables may then be selected by an appropriate multiple testing procedure.

\subsection{Outline}
This remainder of the paper is organized as follows. In the next section, we describe in detail the three main steps of the DCM procedure. Section \ref{theory} provides a closer examination of the test statistic used in the procedure, including a discussion of its asymptotic distribution. We apply DCM to simulated data in Section \ref{sims}, and compare the results to possible alternative procedures based upon existing work. Finally, in Section \ref{realdata} we present the results of two applications of DCM, to the aforementioned TCGA dataset and to brain activity data from the multi-institutional Human Connectome Project.

\section{The DCM Procedure} 
\label{algorithm}

In this section, we present details of the three components of the proposed DCM procedure: initialization, set update, and residualization. The initialization step employs a simple greedy algorithm to select an initial variable set $A$. Once the initial set is determined, it is passed to an update algorithm that iteratively refines the set, making use of a hypothesis testing framework to test variables for differential correlation. When an estimated DC clique is found, the residualization step prepares the data for further search by removing the differential correlation of the discovered set.  

The DCM procedure is summarized below. For detailed pseudocode, see Appendix \ref{pseudocode}.

\begin{samepage}

\begin{framed}

\begin{centering}

\underline{\sc{The DCM Procedure}}

\end{centering}

\begin{enumerate}

\item[$\rhd$] \emph{Initialization:}  Identify a good initial variable set $A$ using a greedy algorithm that identifies a local maximum
of a simple score function.
	
\item[$\rhd$] \emph{Iteration:}  Refine the initial set $A$. At each iterative step, repeat the following until termination.

\begin{enumerate}
	\item[$\rhd$] \emph{Test} the differential correlation of each variable $i$ with respect to $A$. 
	Let $A'$ be the set of variables with significant differential correlation, as determined by 
	and FDR controlling multiple testing procedure.
	\item [$\rhd$] \emph{Terminate} if $A' = A$ or a cycle is observed.
	\item[$\rhd$] \emph{Update:}  Set $A$ to be $A'$.
	
\end{enumerate}

\item[$\rhd$] \emph{Return:} Output variable set $A$.
\item[$\rhd$] \emph{Residualization:}  Remove the effect of the DC clique $A$ and repeat search with new initial set.

\end{enumerate}

\end{framed}

\end{samepage}

We now provide a more in-depth discussion of each step of the procedure.

\subsection{Notation}
\label{sec:notation}
We assume that the data under condition 1 consists of $n_1$ independent samples drawn from 
a distribution $F_1$ with correlation matrix $\RR_1$, and that the data under condition 2 
consists of $n_2$ independent samples drawn from a distribution $F_2$ with correlation 
matrix $\RR_2$. Let $\mathbb{X}_1 = (\UU_1, ..., \UU_{p}) \in \reals^{n_1 \times p}$ and 
$\mathbb{X}_2 = (\VV_1, ..., \VV_p) \in \reals^{n_2 \times p}$ denote the resulting data
matrices in standard sample-by-variable form. Thus $\UU_j \in \reals^{n_1}$ denotes the 
measurements of variable $j$ under condition 1, while $\VV_j \in \reals^{n_2}$ denotes 
the measurements of variable $j$ under condition 2. 
Let $\mathbb{X}_{1, A} = (\UU_j)_{j \in A}$ and $\mathbb{X}_{2, A} = (\VV_j)_{j \in A}$ denote
the restriction of $\mathbb{X}_{1}$ and $\mathbb{X}_{2}$, respectively, to a variable
set $A \subset [p]$. Similarly, let $\RR_{1,A}$ and $\RR_{2,A}$ denote the correlation matrices of the restricted datasets. 

Let $\tilde{\UU}_j$ and $\tilde{\VV}_j$ be the standardized versions of
$\UU_j$ and $\VV_j$, respectively, and define 
$\tilde{\mathbb{X}}_1 = (\tilde{\UU}_1, ..., \tilde{\UU}_{p})$ and 
$\tilde{\mathbb{X}}_2 = (\tilde{\VV}_1, ...,\tilde{\VV}_p)$.
Finally, let $\widehat{\RR}_1$ and $\widehat{\RR}_2$ denote the usual sample correlation matrices 
of $\mathbb{X}_1$ and $\mathbb{X}_2$, respectively (and $\widehat{\RR}_{1,A}$ and $\widehat{\RR}_{2,A}$ those of the appropriate restricted datasets). Thus 
$
\big( \widehat{\RR}_1 \big)_{ij} 
\, = \, 
\corh{\UU_i, \UU_j}
\, = \,
\big( \tilde{\mathbb{X}}_1^T \tilde{\mathbb{X}}_1 \big)_{ij}
$
and a similar relation holds for $\widehat{\RR}_2$. 

\subsection{Initialization}
\label{sec:init}

The set update procedure in the second step of DCM readily identifies 
variables that are significantly differentially correlated relative to a given 
variable set $A$, and is most effective when the initial set of variables exhibits
at least low levels of differential correlation. (When applied to a randomly
chosen set of variables, the set update procedure typically returns an empty set.)
The core search procedure could be run exhaustively, beginning with every 
variable set $A \subset [p]$, but this is not computationally feasible for data sets
of high or moderate dimension. As an alternative, we identify initial variable sets
exhibiting a moderate degree of differential expression using a greedy search
procedure. We then can pass this initial skeleton clique to the set update process to be 
fleshed out into a final estimated DC clique. 

The initialization procedure seeks a local maximum of the score function
\[
S(A) = \dsum_{ i,j \in A } \left\{ ( n_1 - 3 )^{1/2} \; \varphi \left( \widehat{\RR}_1 \right) - ( n_2 - 3 )^{1/2} \;\varphi \left( \widehat{\RR}_2 \right) \right\}_{ij}
\]
where $\varphi$ is the element-wise Fisher transformation of sample correlations, namely
\[
\varphi(r) =\frac{1}{2} \log \left(\frac{1-r}{1+r}\right).
\]

The Fisher transformation and subsequent weighting by degrees of freedom ensures that the first and second terms in the sum have approximately equal variances for each $i,j$. In this way, we account for the natural variance of the individual elements of $\widehat{\RR}_1$ and $\widehat{\RR}_2$ as well as possible imbalance in sample sizes across the two datasets.

To find a local maximizer of $S(\cdot)$, we begin with a random seed $A$. We consider only pairwise swaps, replacing an element of $A$ with one from $A^c$; the set $A$ is then updated by making the swap that produced the largest increase in the score. Since exactly one element is added and removed at each stage, the size of the variable set remains constant. The cardinality of $A$ is user-specified (with a default of 50). Due to the subsequent set update procedure, we find that a wide range of initial cardinalities result in the same final outcome.\footnote{As a rule, erring on the side of initial cardinalities that are \emph{smaller} than the expected output set size is advisable, to avoid drowning out signal with too much noise.} Because of the random seeding, the algorithm is not purely deterministic; however, in practice the same local maximum is reached from most seeds.

Further discussion of the initializing algorithm is available as supplemental material; pseudocode for its implementation may be found in Appendix \ref{pseudocode}. A closely related method is implemented in Section \ref{sims} for comparison with DCM.

\subsection{Core set update procedure}
\label{multtest}
\label{sec:mult}

The heart of the DCM procedure is the set update algorithm, which makes use of multiple testing principles to iteratively refine a variable set $A$. Recall that the goal of DCM is to estimate DC cliques from the data. To this end, the set update procedure is designed to identify variable sets exhibiting the properties of a true DC clique up to a level of statistical significance.

Consider a single iterative step, at which we update a given variable set $A$. We wish to determine whether each variable $i$ (including those in $A$ itself) ought to be included in the updated set $A'$. Since our eventual goal is to discover a DC clique, we perform hypothesis tests based upon the principles of Definition \ref{def:DCc}. For fixed $A$, the tests for variable $i$ may be written as 
 \begin{equation} \label{eq:hypo}
H_0 (i, A): \Delta(i,A) = 0 \tab \tab  \text{vs.} \tab \tab H_a (i, A): \Delta(i,A) > 0 \, .
\end{equation}
Recall that $\Delta(i,A)$, as defined in \eqref{eq:delta}, is a difference of 
average pairwise correlations between $i$ and elements of $A$ from $\RR_1$ and $\RR_2$, so this amounts to a test of differential correlation \emph{relative} to a fixed set $A$. Our updated set is then given by $$A' = \{i \st H_0(i, A) \text{ was rejected by simultaneous multiple hypothesis testing}\}.$$ 
In order to test the hypotheses in \eqref{eq:hypo}, we require a test statistic. A natural choice is the corresponding sample quantity,
\begin{equation}
\label{eq:test}
\hat{\Delta}(i,A) = \frac{1}{ |A| } \dsum_{ j \in A } ( \widehat{\RR}_1 - \widehat{\RR}_2 )_{ij}.
\end{equation}
In addition to being a straightforward choice, this test statistic exhibits several desirable properties discussed in Section \ref{theory}. 

Let $\delta(i,A)$ denote the realized value of the test statistic for a particular dataset. It is clear that large positive values of $\delta( i,A)$ provide support for the alternate hypothesis in \eqref{eq:hypo}, while values that are negative or close to zero provide evidence in favor of the null. 
Thus, in order to test the hypotheses, we calculate a p-value of the form
\[
\p(i:A) = \mathbb{P}_0 \left( \hat{\Delta}(i,A) > \delta(i,A) \right) \, ,
\]
where the probability $\mathbb{P}_0$ is the (unknown) distribution of $\hat{\Delta}(i,A)$ under 
the null hypothesis $\Delta(i,A) = 0$. 
Since we make no assumptions about the distributions of data ($F_1$ and $F_2$), we instead make use of asymptotic results to approximate the above probability. 
We show in Section \ref{sec:distn} that, under appropriate regularity assumptions, and for large enough sample sizes $n_1, n_2$,  
\begin{equation}
	\p(i:A) \approx 1-  \Phi \left( \frac{ \delta(i,A) }{ \hat{\sigma}_0 } \right) ,
	\label{eq:pval}
\end{equation}
where $\hat{\sigma}_0^2$ is an estimate of the variance of $\hat{\Delta}(i,A)$.

The collection of p-values $\{ \p(i:A) \}_{i = 1}^p$ measure the significance of the differential correlation of each variable relative to $A$. In order to select a set of significant variables $A'$, we apply the modified FDR procedure of \citeauthor{bhy} to the p-values. Specifically, we carry out the following steps

\begin{enumerate}
\item Order the p-values $\{ \p(i:A) \}_{i = 1}^p$ as $\{ \p_{(1)} , ..., \p_{(p)}\}$.
\item Define the cutoff index $k^*$ by
\[
k^* = \max \left\{ k \st  \p_{(k)} < \left( \dsum_{i = 1}^k 1/i  \right)^{-1} \left( \frac{k \alpha}{p} \right) \right\}.
\]
\item Let $A' = \{ i \st \p(i:A) \leq \p_{(k^*)} \}$.
\end{enumerate}

Recall that we impose no assumptions on the structure of correlation matrices $\RR_1$ and $\RR_2$. 
In particular, it is possible that p-values $\p(i:A)$ and $\p(j:A)$ may 
be negatively correlated, and negative correlation is common in real data sets.  
Most common multiple testing methods assume independence or positive dependency between p-values.
The possibility of negative dependency of p-values necessitates the more conservative multiple testing method of \citet{bhy},
which controls the expected False Discovery Rate at level $\alpha$.

The main search procedure terminates when it degenerates ($A = \emptyset$) or converges ($A = A' \neq \emptyset$).  For the degenerate case, the interpretation is simple: the initial set (chosen in the first step of the DCM procedure) was not sufficiently differentially correlated to yield an interesting result.  In the second case, we have identified an empirical DC clique, in the sense that by design, a nonempty fixed point $A$ meets the first requirement of a DC clique in Definition \ref{def:DCc} up to a level of statistical significance.  The only other possible outcome of the iterative search procedure is a multi-set cycle; this corner case is discussed in Section \ref{sec:special}. \\

\noindent {\bf Remark.} Iterative updating using multiple testing was applied by \citet{essc} in the context of 
community detection for binary networks.  DCM employs a similar search structure in the context of differential correlation. 
Whereas the method of \citeauthor{essc} relies on a null network distribution that is based on observed node degrees, 
here we make use of a central limit theorem that does not rely on the structural properties of the observations.

\subsection{Residualization}
\label{resid}

In general, we expect multiple DC cliques to be present in data. The residualization step allows the DCM procedure to search the same dataset many times, avoiding repeat identification of prior results. This is accomplished by generating new residualized data matrices $\mathbb{X}_1', \mathbb{X}_2'$ after each (non-degenerate) termination of the set update algorithm. 

For clarity, let us restrict our attention to $\mathbb{X}_1$ (the same process is applied, separately, to $\mathbb{X}_2$). Suppose the set update procedure converges to a set $A$, with $|A| = k > 0$.  We model the correlation matrix of the variables in $A$ as the combination of a rank-one shared group correlation, characterized by $\Lambda \in \reals^{k}$, and an underlying residual 
structure $\Omega \in \reals^{k \times k}$, so that
$\RR_{1,A} 
\, = \,
\Lambda \Lambda^t + \Omega \, .$
The problem of ``factor analysis", or representing correlation matrices by the best lower-rank approximation, is well studied \citep{factora}, and efficient methods for estimating $\Lambda$ are readily available.\footnote{In the DCM software, we use the implementation of \citet{famt} for the R Statistical Software version and the method of \citet{Bishop} for the Matlab version.}   Given an estimate $\hat{\Lambda}$, we generate a new submatrix $\mathbb{X}_{1,A}'$ such that
\[
 \corh{\mathbb{X}_{1,A}'} 
 \, = \,
 \widehat{\RR}_{1, A} - \hat{\Lambda} \hat{\Lambda}^t
 \, = \,
 \widehat{\Omega}.
 \]

We then build $\mathbb{X}_1'$ by replacing $\mathbb{X}_{1,A}$ with the residualized data $\mathbb{X}_{1,A}'$ and leaving the rest of the data matrix unaltered. In this way, neither $\mathbb{X}_1'$ nor $\mathbb{X}_2'$ contain the groupwise structure of $A$ captured by $\Lambda$. 
We are then free to apply the DCM procedure to $\mathbb{X}_1'$ and $\mathbb{X}_2'$ and search for secondary empirical DC cliques.

\subsection{Special Cases}
\label{sec:special}

\emph{Minimality:} A nonempty fixed point $A$ of the set update procedure has the property that, analogously to Definition \ref{def:DCc}, $H_0(i,A)$ is rejected if and only if $i \in A$.  The second condition of Definition \ref{def:DCc}, however, is not guaranteed in general.  It is possible that DCM may select a large set that in truth consists of two (or more) disjoint population DC cliques.  These cases are well addressed by the residualization step.  When a conglomerate estimated DC clique is residualized, the \emph{joint} structure is removed, leaving behind the individual structure of the disjoint DC cliques.  Further runs of the DCM algorithm are then able to identify the separate DC cliques.  

In extreme cases, the sampled data may be such that the disjoint DC cliques are, by chance, correlated enough to have negligible remaining individual structure after residualization.  This renders the multiple DC cliques indistinguishable in the data from a combined DC clique.\\

\noindent \emph{Cycles:} Under certain conditions, the main search procedure may be caught in a cycle of two or more sets, so that termination to a fixed point never occurs.  When the set update procedure oscillates between two sets $A_1$ and $A_2$, we restart the search on the intersection $A = A_1 \cap A_2$. Usually, this results in true convergence to a fixed point in the vicinity of the intersection. On occasion, the same oscillation ($A_1$ to $A_2$) re-emerges, in which case the overlap $A = A_1 \cap A_2$ is selected as the final output. For this overlap set, $H_0(i, A)$ will be rejected for all $i \in A_1, A_2$. It is not quite a fixed point; nevertheless, it shares many properties of a true DC clique, and we consider it worthy of attention.

Cycles of length greater than two are rarely observed in real or simulated data. However, to protect against longer cycles leading to infinite loops, the algorithm is built with an adjustable maximum iteration limit.

\section{Properties of the Test Statistic}
\label{theory}

We now discuss some properties of the test statistic $\hat{\Delta}(i,A)$ used in the calculation of p-values for the set update procedure. 

\subsection{Geometric Interpretation}

The equation for $\dia$ given in (\ref{eq:test}) expresses the test statistic directly in terms of average differential correlation. However, we may also write $\dia$ in an alternate form that yields an informative geometric interpretation.  Let $\tilde{\UU}_i \in \reals^{n_1}$ and $\tilde{\VV}_i  \in \reals^{n_2}$ be the standardized 
measurements of variable $i$ under sample conditions 1 and 2, respectively; and let
\begin{equation}
\label{eq:centroids}
\WW_{1} \, : = \, \frac{1}{|A|} \sum_{j \in A} \tilde{\UU}_j
\ \mbox{ and } \ 
\WW_{2} \, : = \, \frac{1}{|A|} \sum_{j \in A} \tilde{\VV}_j
\end{equation}
be the vector means of the standardized measurements of the variables in $A$ under each condition. It is easily shown that
\[
\frac{1}{|A|} \dsum_{j \in A} \corh{ \UU_i, \UU_j }
\, = \,
\WW_1^{\, t} \tilde{\UU}_i
\, = \,
\|  \WW_{1}  \|  \, \corh{ \tilde{\UU}_{i} , \WW_{1} }  \, .
\]
Thus,
\[
\dia \ = \ \|  \WW_{1}  \|  \, \corh{ \WW_{1}, \tilde{\UU}_{i} } 
\, - \, 
\| \WW_{2}  \|  \, \corh{ \WW_{2}, \tilde{\VV}_{i} } .
\]

Note that the vector $\tilde{\UU}_i$ and the vectors $\{ \tilde{\UU}_j : j \in A \}$ lie on the surface of an 
$(n_1 -2)$-dimensional sphere in $\reals^{n_1}$, and that $\WW_{1}$ is the geometric center
(centroid) of the latter collection. The norm $|| \WW_{1} ||$ is between $0$ and $1$; large
values of $|| \WW_{1} ||$ place the centroid closer to the surface of the sphere, indicating that the
vectors $\{ \tilde{\UU}_j : j \in A \}$ are tightly clustered, or equivalently,
highly intercorrelated. Thus
the quantity $\| \WW_{1} \| \, \corh{\WW_{1}, \tilde{\UU}_{i}}$ weights the similarity of 
$\UU_{i}$ and the centroid $\WW_{1}$ according to the overall similarity of the vectors
$\{ \tilde{\UU}_j : j \in A \}$. Similar remarks apply to $\{ \tilde{\VV}_j : j \in A \}$ and $\WW_{2}$.
The statistic $\dia$ is the difference of the summary measures in conditions 1 and 2.

\begin{figure}[H]
\centering
\includegraphics[width = \textwidth]{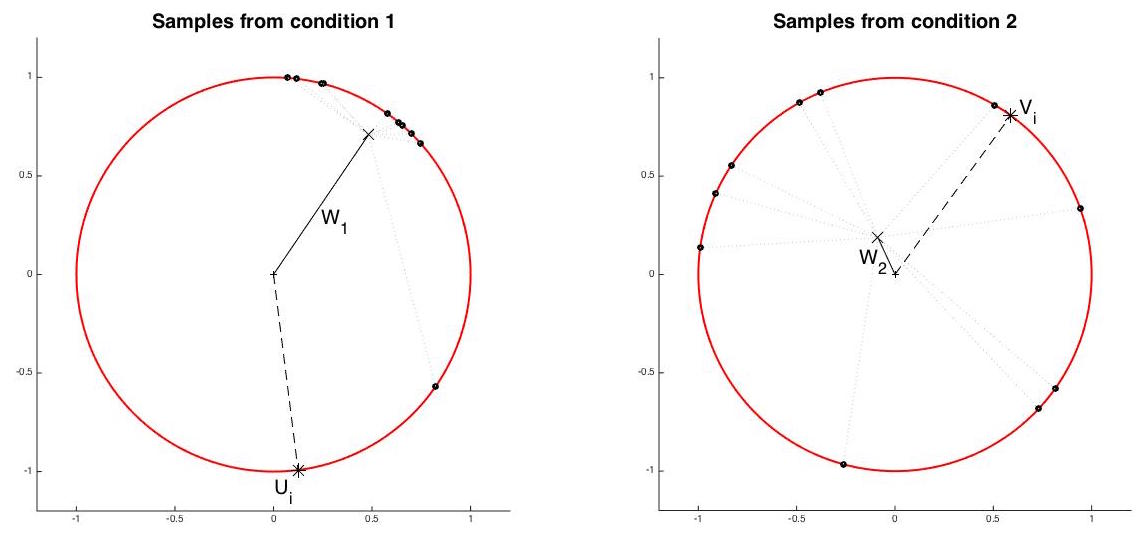}
\caption{Geometric representation of data in two dimensions. Unlabelled points represent the standardized data in group $A$.}
\label{fig:circles}
\end{figure}

Figure \ref{fig:circles} gives a simple two-dimensional representation of 
the geometric picture discussed above. In condition 1, $\UU_{i}$ is not strongly correlated 
with $\WW_1$, but $\| \WW_{1} \|$ is large because the vectors indexed by $A$ are tightly clustered. 
In condition 2, $\VV_{i}$ is strongly correlated with $\WW_2$, but $\| \WW_{2} \|$ is small 
because the vectors indexed by $A$ are not tightly clustered. In this example, 
$\dia$ is close to zero, and we would likely conclude no differential correlation is present.

\subsection{Asymptotic distribution of $\dia$}
\label{sec:distn}

We now discuss the asymptotic distribution of $\dia$, from which the p-values
 used in Section \ref{sec:mult} are derived. First, we make note of a classical result concerning sample correlations.

\begin{theorem}
\label{thm:rp}
\emph{{\bf \citep{steiger2}}} Let $\RR$ be a $p \times p$ correlation matrix, and $\widehat{\RR}$ the corresponding sample correlation matrix based on $n$ i.i.d. samples of $p$-variate data with finite 4th moment. Let $\PP$ and $\widehat{\PP}$ be the vectorized versions of the matrices, of dimension $p^2 \times 1$. 
Then, as $n$ tends to infinity
\[
\sqrt{n} \left( \widehat{\PP} - \PP \right)
\, \Rightarrow \,
\mathcal{N}_{p^2} \left( {\bf 0}, {\bf \Sigma} \right)
\]
where ${\bf \Sigma}$ is a $p^2 \times p^2$ covariance matrix for which a general form is given equations (3.1-3.5) in \citet{southafrica}.
\end{theorem}

Using Theorem \ref{thm:rp} one may evaluate the asymptotic distribution of $\dia$, which
is a function of $\PP$ and $\widehat{\PP}$.

\begin{corollary}
\label{cor:DCLT}
Let $A$ be a fixed index set and let $\dia$ be defined as in \eqref{eq:test}, 
with sample correlation matrices $\widehat{\RR}_1$ and $\widehat{\RR}_2$ based on $n_1$ and 
$n_2$ independent samples from distributions $F_1$ and $F_2$ respectively.  Let $\sigma^2_0 (i,A) := \var{ \dia \given H_0 }$,
where $H_0$ is the null hypothesis in (\ref{eq:hypo}).  Then, under $H_0$,
\begin{equation}
\label{eq:asym}
	\frac{\dia}{ \sigma_0(i, A) }
	\, \Rightarrow \,
	 \mathcal{N} \left( 0, 1 \right)
\end{equation}
 as $\min( n_1, n_2 ) \to \infty$.

\end{corollary}

\noindent {\bf Proof:} For clarity, we first examine only one ``half" of $\dia$. Let

\begin{equation}
\label{hia}
\bar{r}_1(i,A) 
\, = \,
\frac{1}{|A|} \sum_{ j \in A } \left( \widehat{\RR}_1 \right)_{ij}
\tab \text{and} \tab
\bar{\rho}_1(i, A) 
\, = \,
\frac{1}{|A|}  \sum_{ j \in A } \left( \RR_1 \right)_{ij} \, .
\end{equation}
Note that $\bar{r}_1(i,A)$ is a linear function of $\widehat{\RR}_1$ and that $\bar{\rho}_1(i,A)$ is the 
same function applied to the population correlation matrix $\RR_1$. It follows from 
Theorem \ref{thm:rp} that 
\[
\sqrt{n_1} \, \left( \frac{ \bar{r}_1(i,A) - \bar{\rho}_1(i, A)}{ \tau_1^2 (i,A) } \right)
\, \Rightarrow \,
 \mathcal{N} \left( 0, 1 \right) \, .
\]
with $\tau_1^2 (i,A) := \var{ \sqrt{n_1} \, \bar{r}_1(i,A) }$, which has a finite limiting value that can be 
expressed as the mean of appropriate elements of the covariance matrix ${\bf \Sigma}$ in the theorem. 
To apply this result for the full test statistic, we note that 
$\dia = \bar{r}_1(i,A) - \bar{r}_2(i,A)$. Samples from $F_1$ are independent of those from $F_2$, 
so $\bar{r}_1(i,A)$ is independent of $\bar{r}_2(i,A)$, and thus $\dia$ is asymptotically normal.

Under the null hypothesis in \eqref{eq:hypo}, $\bar{\rho}_1(i, A) = \bar{\rho}_2(i, A)$,
and therefore the mean of the limiting distribution of $\dia$ is 0.  The variance of $\dia$ can be 
expressed as the weighted sum 
 \[
	\sigma^2_0(i,A)
	\, = \,
	\frac{\tau_1^2(i,A)}{n_1} + \frac{\tau_2^2(i,A)}{n_2}  \, .
\]
\qed

\vskip.1in

In practice, the variances $\tau_1^2 (i,A) $ and $\tau_2^2 (i,A) $ are not known.  
Thus, we make use of the results in  \citet{steiger2} to identify the asymptotic variance of $\bar{r}_1(i,A)$, 
and a consistent estimator $\hat{\tau}_1$.

\vskip.1in

\begin{lemma}
\label{lem:varest}
Let $r_{ij}$ be the sample correlation of $\UU_i$ and $\UU_j$, and let $r_A := \frac{1}{|A|} \sum_{j \in A} r_{ij}$. For $\ell = 1, 2, ... , n_1$ let 
\[
W_\ell \, : = \, \frac{1}{|A|} \sum_{j \in A} \tilde{\UU}_{j \ell} \, , \tab \text{and} \tab Y_\ell \, : = \, \frac{1}{|A|} \sum_{j \in A} r_{ij} \tilde{U}_{j \ell}^2 \, .
\]
and define
\[
\hat{\tau}_1 
\, = \,
\frac{1}{n_1} \sum_{\ell = 1}^{n_1}
\left\{ \frac{ r_A^2 }{4} \, \tilde{U}_{i \ell}^4
\, - \,
 r_A W_\ell \, \tilde{U}_{i \ell}^3
  \, + \,
\left( \frac{ r_A Y_\ell }{2} \, + \, W_\ell^2 \right) \tilde{U}_{i \ell}^2
 \, - \,
 W_\ell Y_\ell \, \tilde{U}_{i \ell}
  \, +  \,
 \frac{ Y_\ell^2 }{4}
\right\} \, .
\]
If $\bar{r}_1(i,A)$ is defined as in $\eqref{hia}$, then under any sampling distribution satisfying the 
assumptions of Theorem \ref{thm:rp}, we have
\[
 \frac{\hat{\tau}_1}{ \tau_1}
\, \stackrel{p}{\to} \,
1 \tab \text{as} \tab n_1 \to \infty \, ,
\]
where $\tau_1 = \var{ \sqrt{n_1} \, \bar{r}_1(i,A) } $.
 
\end{lemma}

\noindent Lemma \ref{lem:varest} is proven in Appendix \ref{app:varest}.  The results of this lemma allow us to make an efficient and straightforward calculation to estimate the variances of $\bar{r}_1(i,A)$ and $\bar{r}_2(i,A)$ for every $i \in [p]$, used in the testing step of the DCM algorithm. Note that regardless of the size of $A$, the derived formula for $\hat{\tau}_1$ involves basic operations on only three $p \times n_1$ vectors: $\WW, \YY$ and $\tilde{\UU}_i$. Such simplification is important, since in practice the estimates $\hat{\tau}_1$ and $\hat{\tau}_2$ must be calculated for \emph{every} 
variable $i \in [p]$ and for multiple iterative steps of $A$.

\begin{remark}
\emph{The results of Corollary \ref{cor:DCLT} and Lemma \ref{lem:varest} apply to variable sets of fixed cardinality ($|A| = k$) as $n_1$ and $n_2$ tend to infinity. In practice, one may encounter variables sets for which $k > n_1, n_2$.  Simulations suggest that the DCM algorithm still identifies DC cliques with high success and controls false discovery in such settings even when the cardinality of $|A|$ greatly exceeds the sample size.}  
\end{remark}

\section{Simulation Study}
\label{sims}

To test the DCM method against possible alternatives, we implemented a simple study of performance on simulated data. We created artificial datasets containing a single DC clique and compared the results of several methods to the known truth. Although the simulated setting is not a perfect representation of real data situations, it readily illustrates the advantages of DCM as opposed to adaptations of existing methods.

\subsection{Methods implemented}

In the absence of comparable methods, we adapted several existing methods to search for DC cliques.  These
adaptations are described below.

\emph{Mining a single correlation matrix (WGCNA).}  We implemented the Weighted Gene Co-Expression Network Analysis (WGCNA) method of \citet{wgcna} for clustering correlation matrices of gene expression data. This method is a hybrid analysis, which mines for clusters (or ``modules") in a single correlation matrix, then tests each module for differential \emph{expression}. Thus, although the WGCNA method involves both differential and second-order elements, it is not designed to search for DC cliques or similar structures.  For the purposes of this simulation study, we applied WGCNA to samples from class 1 only.  We then tested the output module for differential correlation via a standard t-test over sample correlations in classes 1 and 2.  In this way, we attempted to only select variable sets exhibiting differential correlation, even though WGCNA does not naturally identify modules with this property.

\emph{Estimation of differential correlation matrices (D-EST and FISH).}  We implemented the method of \citet{caizhang} to estimate $\DD = \RR_1 - \RR_2$ by the suggested estimator $\widehat{\DD}$. To search for the seeded DC clique, we converted $\widehat{\DD}$ to a $[0,1]$ scale (i.e. a dissimilarity matrix $\DD^*$), then performed hierarchical clustering on $\DD^*$.  We  extracted the first cluster in the dendrogram, starting at the bottom, of size less than or equal to the size of the seeded DC clique.  This approach necessarily returns a nonempty variable set of approximately correct size; in non-simulated settings, one would not know the size of true DC cliques.

In a similar fashion, we calculated
\[
\widehat{\DD}_{Fisher} = (n_1 - 3)^{1/2} \;\varphi\left(\widehat{\RR}_1\right) - 
(n_2 - 3)^{1/2}\;\varphi\left(\widehat{\RR}_2 \right)
\]
where $\varphi$ is the (element-wise) Fisher transformation of sample correlations.  Note that $\widehat{\DD}_{Fisher}$ is not, in fact, an estimator of $\widehat{\RR}_1 - \widehat{\RR}_2$, but it is analogous to $\widehat{\DD}$ in that large entries in general correspond to large correlation differences.  As with $\widehat{\DD}$, we converted $\widehat{\DD}_{Fisher}$ to a dissimilarity matrix and extracted a cluster from a hierarchical dendrogram.

\emph{Detection of isolated changes in correlation structure (DCP).}  We applied the Differential Correlation Profile (DCP) method of \citet{dcgl}. This method takes a permutation approach to variable selection. Each variable $i \in [p]$ is assigned a measure $\Psi_i$ of overall differential correlation. The samples are then randomly permuted many times, each time calculating new measures $\Psi_i'$, and significant variables are chosen by comparing $\Psi_i$ to the permutated values. This approach identifies a list of individual differentially correlated variables, rather than a united set; for the purposes of this study, we considered the collection of selected variables to be the estimated DC clique.

\emph{Alternate testing step (CLASSIC).}  Finally, we implemented an altered version of DCM that uses the same algorithmic framework but makes use of classical results to test for differential correlation.  
Instead of testing $H_0(i, A): \dia = 0$ in the DCM algorithm (as in Section \ref{sec:mult}), we test $H_0(i,A): R_1(i,A) = R_2(i,A)$, or equality of the vectors of correlations between $i$ and elements in $A$.  We adapted the testing method of \citet{classic} for a one-sided alternate hypothesis; namely, $H_a(i,A): \max \{ R_1(i,A)  \} > \max \{ R_2(i,A) \}$.

\subsection{Simulated Data}

We simulated data with a single embedded DC clique, consistent with Definition \ref{def:DCc}. Our study varied the following parameters: size of the DC clique ($k$), total number of variables ($p$), strength of the correlations ($\rho_1$ and $\rho_2$), samples sizes of the two datasets ($n_1$ and $n_2$), and background data. We considered three background types: uncorrelated, positive, and real data. To elaborate, let $\SSS_1$ and $\SSS_2$ be matrices that are 1 on the diagonal, $\rho_1$ or $\rho_2$ (respectively) on the off-diagonal for indices 1 to $k$, and 0 otherwise. Then, the three types of data simulations were:

	\begin{itemize}
		\item[] \emph{Uncorrelated Gaussian.} The data was generated from multivariate Normal distributions with correlations $\RR_1 = \SSS_1, \RR_2 = \SSS_2$.
		\item[] \emph{Positively correlated Gaussian.}  The correlation matrices of the data were $\RR_1 = \SSS_1 + \rho_1/3$, $\RR_2 = \SSS_2 + \rho_1/3$. 
		That is, the correlation matrix was boosted everywhere (except on the diagonals) in equal part for both datasets.
		\item[] \emph{Real data.}  The dataset was derived from a real-world data source\footnote{The Cancer Genome Atlas; Luminal A tumor subtype} with samples randomly assigned one of two sample classes. By adding a common vector to the first $k$ rows of the data matrix, we induced further correlation $\rho_1$ and $\rho_2$ in part the data.
	\end{itemize}

We found that all methods behaved similarly with regard to changes in sample sizes $n_1, n_2$ and clique size $k$ (relative to $p$). Here, we present only the results regarding $\rho_1$ vs. $\rho_2$ and different background types, to illustrate key differences in performance between methods. By default, the other parameters were set to be $n_1 = n_2 = 100$, $k = 100$, and $p = 1000$. 

\subsection{Results}

We applied the five proposed methods (DCM, WGCNA, D-EST, FISH, DCP, and CLASSIC) to a variety of values of $\rho_1$ and $\rho_2$ for each of the three background data types. For each $(\rho_1, \rho_2)$, we ran 100 random trials and averaged the results.  The success of the methods was measured by the false positive rate (FPR), the percentage of variables in a selected set that were not in the seeded DC clique, and the true discovery rate (TDR), the percentage of detected variables from the true DC clique. That is, if $B$ was the output variable set of a procedure and $A = (1, \ldots, k)$ was the embedded DC clique, then
\[
\text{False Positive Rate} = \frac{\left| B \, \backslash \, A \right|}{|B|} 
\tab \text{ and } \tab
\text{False Negative Rate} = \frac{ \left| A \, \backslash \, B \right|}{\left|A\right|}.
\]

\begin{figure}[H]
\centering

\includegraphics[width = \textwidth]{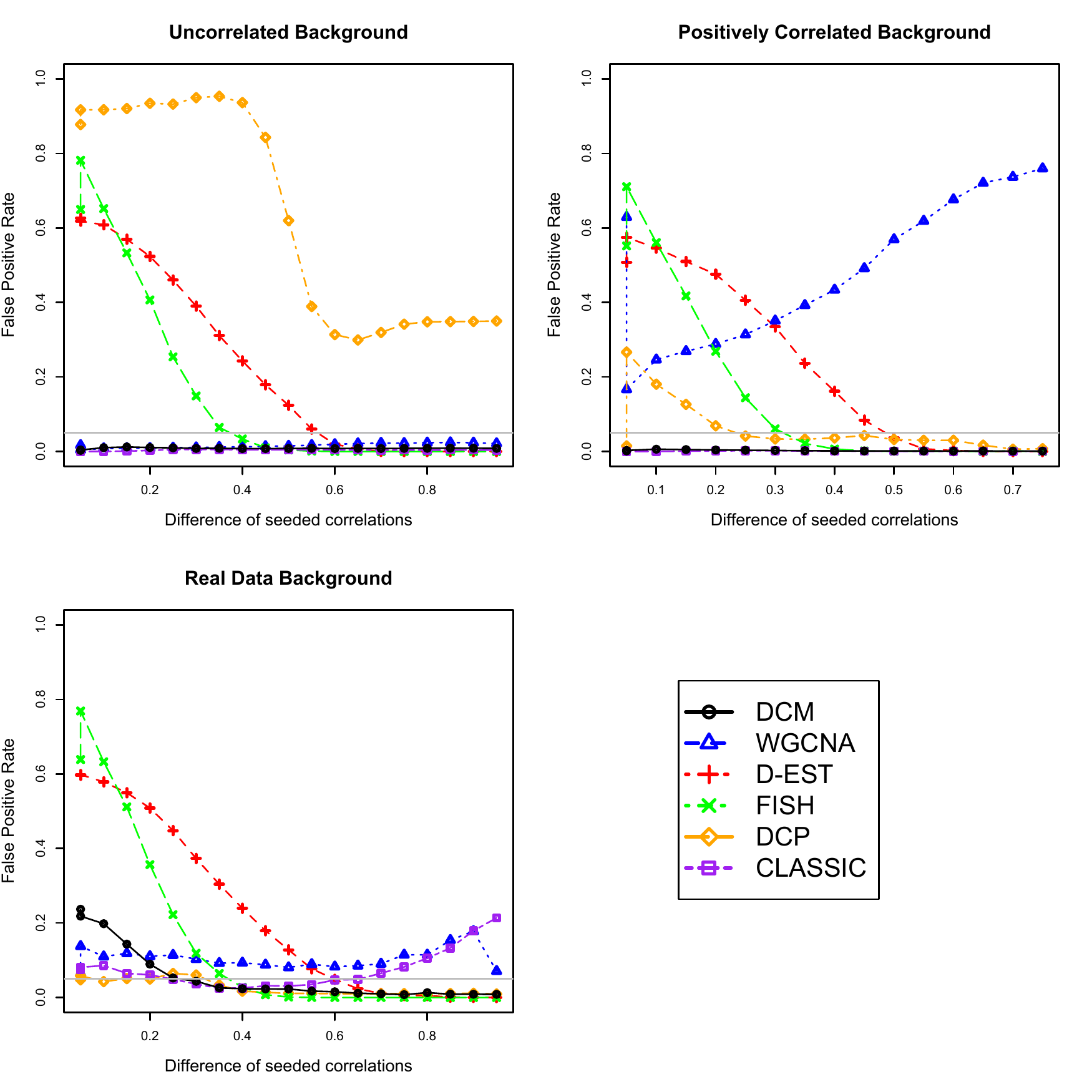}

\caption{Average False Positive rate for each data type. ($n1 = n2 = 100, k = 100, p = 1000$).}

\label{fig:type1}
\end{figure}

\begin{figure}[H]
\centering

\includegraphics[width = \textwidth]{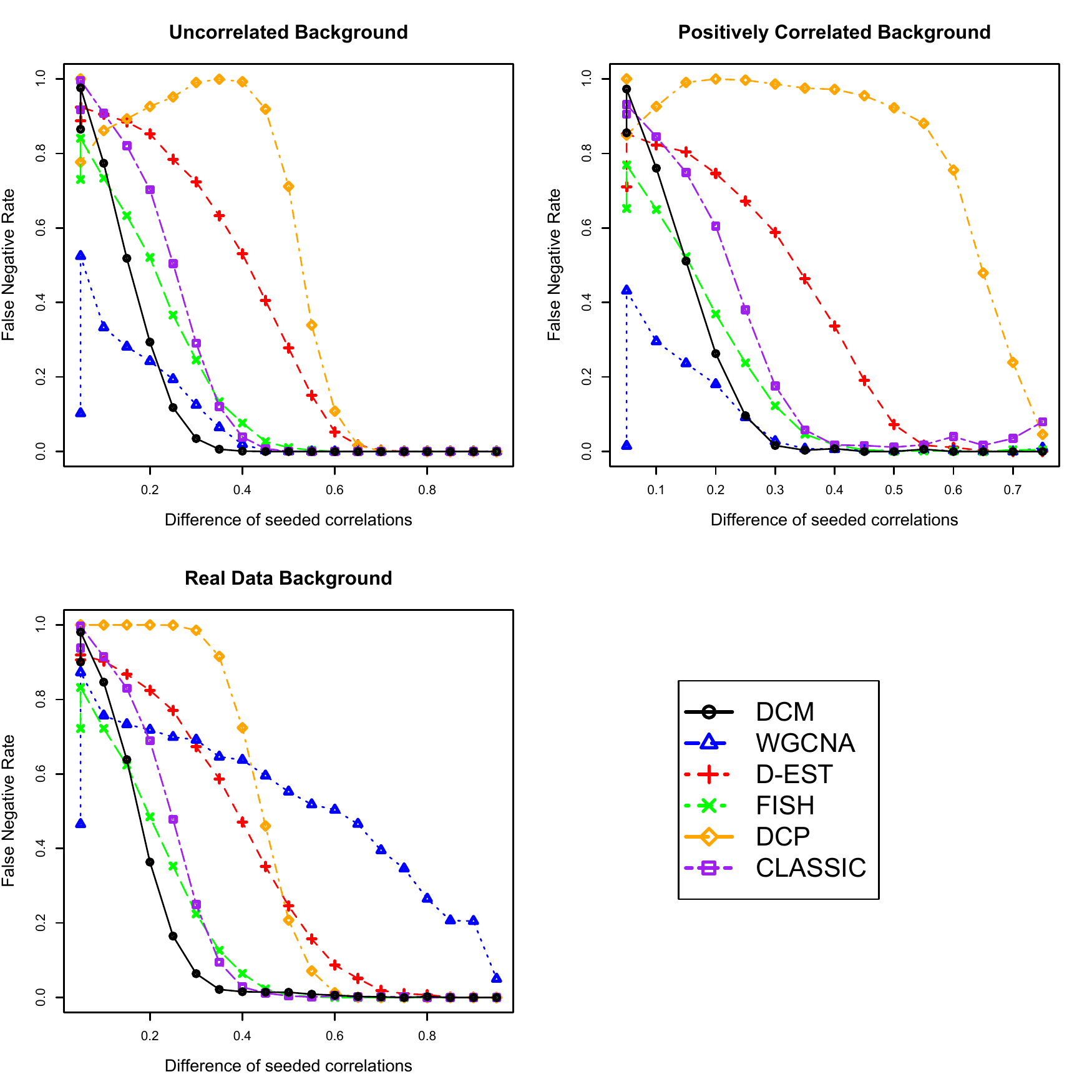}

\caption{Average False Negative rate for each data type. ($n1 = n2 = 100, k = 100, p = 1000$).}

\label{fig:type2}
\end{figure}

Figure ~\ref{fig:type1} shows the False Positive rate for each method as the seeded differential correlation gets stronger ($\rho_1 - \rho_2$ grows), with other parameters fixed at default values, for each background type.  Figure \ref{fig:type2} shows the same for False Negative rate.  In both cases, values close to zero are desirable, as they represent low error in identifying the true seeded DC clique. 

We note that DCM controls the rate of false positives in all cases, except for some error when there is very low signal size in the real data case, which may be due to actual signal being present in the randomized real data.  DCM also begins to reliably detect DC cliques at a lower signal size - around a correlation difference of 0.2 - in every setting except for the simple uncorrelated background setting, where WGCNA has slightly better performance.  When the background is highly correlated, but not differentially correlated, WGCNA is not able to control the false positive rate.  This is because WGCNA is designed to be a single-dataset mining method. When the background data shows strong correlation, WGCNA correctly identifies a large correlated module in the first dataset. Even if these modules could be tested for differential correlation, there is no method inherent to WGCNA for extracting the DC clique from surrounding correlated (but not differentially correlated) data.

The D-EST and FISH methods behave as expected; because our approach necessarily returns a nonempty variable set, the false positive rate is high for small signal.  However, even if the false positives were perfectly controlled in some way, these methods show a higher false negative rate than DCM.  Similarly, the adjusted DCM algorithm with the classical test (CLASSIC) controls error, but is less powerful than our method.

Finally, the DCP method vastly overselects variables in the uncorrelated background case. This is likely because the mutual behavior of the variables in $A$ induces some correlation structure in the background variables; Figure \ref{fig:example} illustrates this phenomenon, as there is some pattern in the cross correlation between variables in $B$ and $A$. This result emphasizes the danger of the common approach of looking for isolated changes in correlation structure of individual variables, rather than searching for DC cliques: vestigial correlation patterns may be misleading. (When the background has structure, as in positively correlated or real data, the induced pattern is overshadowed, so the high false positive rate of DCP does not occur.)

\begin{figure}[H]
\centering

\includegraphics[width = \textwidth]{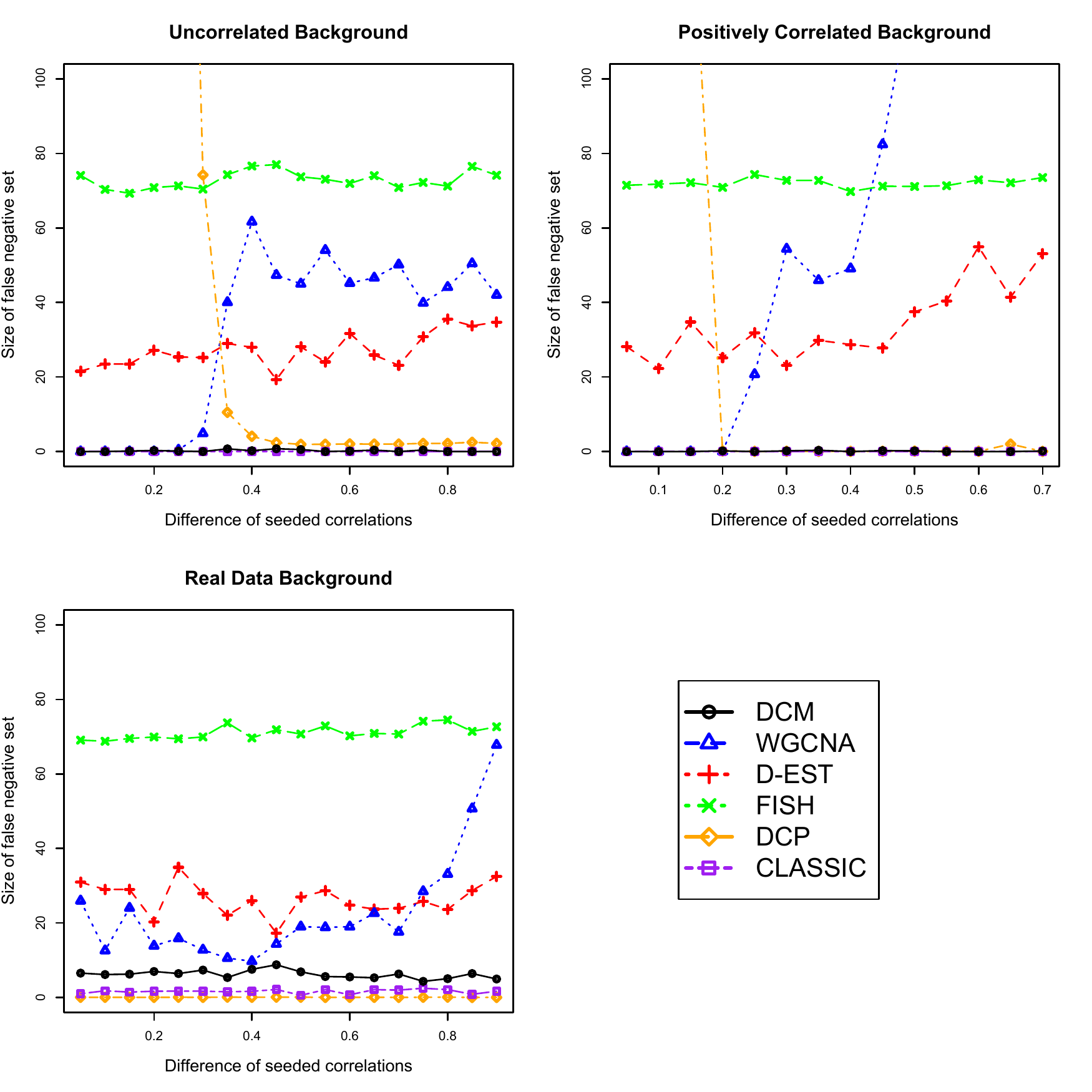}

\caption{Average size of set when no DC clique is present, for each data type. ($n1 = n2 = 100, k = 100, p = 1000$).}

\label{fig:eqrho}
\end{figure}

We wish to note in particular the behavior of these methods when the correlation difference is zero.  These simulations represent settings where correlation is present in both datasets, but the structure is identical across datasets rather than differential.  We expect such scenarios to arise commonly in real data, where groups of variables may be universally correlated without regard to the particular sample conditions being studied. It is important that a search procedure does not identify nonexistent DC cliques in these situations.  Figure \ref{fig:eqrho} shows the average size of selected sets as a function of the strength of the \emph{non-differential} correlation.  Values at zero represent cases where the method successfully avoided identifying the misleading correlation as a DC clique; large values represent the discovery of a false positive set.

It is clear that all methods except DCM and CLASSIC are prone to error when non-differential correlation is present. This striking difference in performance illustrates the ways in which existing methods are simply not designed for the specific case of DC cliques. Only DCM-type algorithms include a testing-based element, which allow them to dismiss observed correlation for which there is not enough evidence of differential behavior.  (In the real data case, DCM sometimes selects small nonempty sets in spite of the lack of seeded DC clique; again, this is likely due to the fact that permuting the samples in real data will sometimes produce signal.)

\subsection{A note about computation}

Of the methods tested, only DCM, CLASSIC, and WGCNA are computationally practical for large $p$ ($\sim 10^5$ or more). The methods FISH and D-EST have memory demands on the order of $p^2$, as they are based upon the estimation of the full $p$ by $p$ dissimilarity matrices, $\widehat{\DD}_{Fisher}$ and $ \widehat{\DD}$.  Permutation-based methods such as DCP are even more infeasible, since they require the computation of a $p$ by $p$ correlation matrix for each of many permutations. 
\section{Data Analyses}
\label{realdata}

\subsection{TCGA}
\label{tcga}

We now expand upon the real data application referenced in Figure \ref{fig:example}. Recall that we applied the DCM procedure to samples from two pre-determined breast cancer subtypes:  Her-2 and Luminal B. A total of 18 empirical DC cliques (more correlated in Her-2 than in Luminal B) were discovered, ranging in size from 13 to 165 genes. 
To illustrate how this information may be useful to genomic research, we briefly discuss one of the discovered gene sets. The set of interest contained 46 genes, listed alphabetically in Table \ref{tab:genes}. These genes are found to be highly associated with immune response, particularly the HLA (Human Leukocyte Antigen) gene class, represented by six of the genes in the set. Researchers are interested in understanding how and why some cancer subtypes trigger immune response while others do not. For example, \citet{iglesia} showed that prognosis was improved for patients with Her-2 and basal-like subtypes showing higher immunoreactive response. Further exploration of DC cliques such as the one in Table \ref{tab:genes} may further understanding of the gene interactions that drive immune response.


\begin{table}[H]
\caption{Genes selected in empirical DC Clique for Her2 vs. Luminal B samples.}
\label{tab:genes}
\centering
\footnotesize{
\begin{tabular}{|ccccccc|}
\hline
&&&&&& \\
AGER & amt & APOL1 & ARPC4 & B2M & BATF2 & BTN3A2  \\
BTN3A3 & C19orf38 & calml4 & CCDC146 & CHKB-CPT1B & echdc1 & ETV7 \\
 EXOSC10 & FBXO6 & GBP1 & GBP4 & GJD3 & gnb3 & {\bf HLA-A} \\
  {\bf HLA-B} & {\bf HLA-C} & {\bf HLA-E} & {\bf HLA-F} & {\bf HLA-H} & HSH2D & IDO1 \\
   IL15 & Irf1 & LOC115110 & LOC400759 & LOC91316 & micB & Myo15b \\
    OASL & PILRB & Rec8 & Rufy4 & SAMD9L & SEC31B & STAT1 \\
     tap1 & Tapbp & TTLL3 & TXNDC6 & Ube2l6 & Zbp1 & \\
     &&&&&&\\
     \hline
 \end{tabular}
 }
 \end{table}

\subsection{The Human Connectome Project}
\label{hcp}

The Human Connectome Project is a multi-institutional venture aimed at mapping functional connections between parts of the human brain. The project has collected vast amounts of brain scan data, all of which is publicly available to researchers online at {www.humanconnectome.org}. In this analysis, we made use of a dataset from the ``500 Subjects MR" data release, which consists of functional magnetic resonance imaging  (fMRI) brain scans for 542 healthy adult subjects. Participants performed a variety of tasks during the MR scan, designed to isolate certain types of brain functionality. Activation levels were recorded over time for $\sim$30,000 voxels (3D coordinate locations in the brain's white matter interior)  and $\sim$60,000 greyordinates (indexed locations over the grey matter brain surface).

In this paper, we applied DCM to data from a single subject.\footnote{Subject \#101006, a 35-year-old female.}  We compared two task categories:
\begin{itemize}
\item[] \emph{Language-based tasks:}  During the scan, subjects were told brief stories and asked to answer questions after each one about what they were told.
\item[] \emph{Motor-based tasks:}  Subjects were attached to motion sensors at the hands, feet, and tongue. They were then asked to move one appendage at a time, in blocks of repetitions.
\end{itemize}

\noindent DCM to searched amongst 91,282 brain locations (or nodes) for DC cliques that exhibit more correlation over time during language tasks than during motor tasks.

\begin{figure}[!h]

\includegraphics[width = 0.45\textwidth]{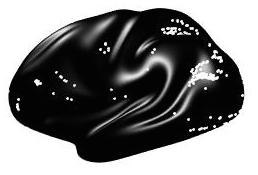}
\includegraphics[width = 0.45\textwidth]{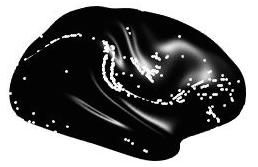}
\caption{Brain locations of DC clique for languages tasks versus motor tasks.}
\label{fig:brains}

\end{figure}

The first empirical DC clique selected by DCM contained 913 nodes located on the cortical surface. These nodes, or ``greyordinates", are visualized as points on the smoothed exterior of the brain in Figure \ref{fig:brains}.  The clear locational pattern in the nodes - despite the fact that the analysis did not take location into account - is striking.  Additionally, the empirical DC clique in Figure \ref{fig:brains} includes a concentrated group in the rear of the left cortex. This general brain region is known to be specifically associated with language processing and auditory input (Wernicke's Area, see \citet{wernicke}).

We also studied two other artifacts of the data for comparison. First, we identified the 1000 nodes exhibiting the strongest differential first-order behavior. These show higher mean activation during the language tasks than during the motor tasks, as measured by standard two-sample t-tests. We saw a clear grouping of nodes in the right frontal lobe. This pattern is unsurprising and appears in many studies of brain functionality that examine differential activation for language processing \citep{rfront}. This basic first-order analysis suggests that differential correlation is not redundant. None of the empirical DC cliques selected by DCM show high frontal lobe concentration; instead, they exhibit ``trail-like" patterns such as the ones shown in Figure \ref{fig:brains}. 

Second, we identified 1000 nodes found to be highly correlated over time for the language task data, irrespective of their behavior in the motor task data. These nodes were observed to be very tightly grouped in the interior left hemisphere.  This is likely due to the nature of data measurement: fMRI brain scans measure oxygen flow in the brain, so measurements for adjacent regions tend to ``blur" and show high artificial correlation \citep{fmri_depend}. In this case, the same node set is also highly correlated during motor tasks, suggesting that it is likely a byproduct of data collection. Even if this node set does represent a meaningful result - regions, perhaps, that are universally correlated regardless of task - it is not differential. 

This example illustrates the advantage of taking a differential approach like DCM. Effects due to fMRI-driven spatial correlation or strong universal correlation can drown out signal that is truly specific to a particular sample condition. By comparing language tasks to the similar but distinct condition of motor tasks, we are able to isolate signals that are unique to language processing.  The fact that the identified DC cliques show emergent locational patterns suggests that DCM is capturing a true facet of the data rather than arbitrary correlation. Since this output is unique in form, while maintaining some consistency with known brain functionality, we believe it merits further scientific investigation.

\section{Conclusion}

In this paper, we have introduced a new statistical method, DCM, to identify differentially correlated variable sets from observed data. The DCM algorithm has been shown to be built on statistical principles with a theoretical basis, and to perform accurately and efficiently in a variety of settings. Unlike existing methods, DCM is specifically built to discover DC cliques, and its underlying testing process controls error. Additionally, the DCM software can be run on extremely high dimensional data without large memory demands or long runtimes. Preliminary data analysis results in the application areas of both gene expression data and brain activation are encouraging.

Code for public use of DCM is freely available at {http://kbodwin.web.unc.edu/software/}.

\section*{Acknowledgements}

The authors wish to thank Yin Xia, Andrey Shabalin, Katherine Hoadley, and Kimberly D.T. Stachenfeld for their contributions.

\begin{appendices}

\section{Variance Estimator}
\label{app:varest}
\subsection{Proof of Lemma \ref{lem:varest}} 

We wish to show that if $\bar{r}_1(i,A)$ is defined as in \eqref{hia}, for fixed $i \in [p]$ and $A \subset [p]$, then
\[
\hat{\tau}_1 
 \, = \,
\frac{1}{n_1} \sum_{\ell = 1}^{n_1}
\left\{ \frac{ r_A^2 }{4} \, \tilde{U}_{i \ell}^4
\, - \,
 r_A W_\ell \, \tilde{U}_{i \ell}^3
  \, + \,
\left( \frac{ r_A Y_\ell }{2} \, + \, W_\ell^2 \right) \tilde{U}_{i \ell}^2
 \, - \,
 W_\ell Y_\ell \, \tilde{U}_{i \ell}
  \, +  \,
 \frac{ Y_\ell^2 }{4}
\right\} \, .
\]
is equal to the consistent estimator for the variance of $\bar{r}_1(i,A)$ given in \citet{steiger2}.  Define
\[
r_{ijkh} \, := \, \sum_{\ell = 1}^{n_1}   \tilde{U}_{i \ell} \tilde{U}_{j \ell}  \tilde{U}_{k \ell} \tilde{U}_{h \ell} \, .
\]
Then, for the variance of an average of sample correlations sample correlations, $\tau_1 \, = \, \frac{1}{|A|^2} \sum_{j,k \in A}\cov{ r_{ij} , r_{ik} }$, equation (5.1) of \citet{steiger2} gives the consistent estimator

\begin{align*}
\hat{\tau}_1
& \, = \,
\frac{1}{|A|^2}  \sum_{j,k \in A} \bigg[ 
r_{iijk}
\, + \,
\frac{1}{4} r_{ij} r_{ik} ( r_{iiii} + r_{iijj} + r_{iikk} + r_{jjkk} ) \\
& \tab \, - \,
\frac{1}{2} r_{ij} ( r_{iiik} + r_{ijjk} )
\, - \,
\frac{1}{2} r_{ik} ( r_{iiij} + r_{ijkk} )
\bigg] \\[.1in]
& \, = \,
\frac{1}{|A|^2} \sum_{j,k \in A} r_{iijk}
\, + \,
\frac{1}{4} r_A^2 r_{iiii}
\, + \,
\frac{1}{2 |A|} \sum_{j \in A} r_{ij} r_{A} r_{iijj} 
\, + \,
\frac{1}{4|A|^2} \sum_{j,k \in A} r_{ij} r_{ik} r_{jjkk} \\
& \tab \, - \,
\frac{1}{|A|} \sum_{j \in A} r_{A} r_{iiij}
\, - \,
\frac{1}{|A|^2} \sum_{j,k \in A} r_{ij} r_{ijjk}
\end{align*}

Using the notation of Section \ref{sec:distn}, we can write
\begin{align*}
&\frac{1}{|A|^2} \sum_{j,k \in A} r_{iijk} 
\, = \, 
\frac{1}{n_1} \sum_{\ell = 1}^{n_1}  \left( \frac{1}{|A|^2} \sum_{j,k \in A} \tilde{U}_{j \ell} \tilde{U}_{k \ell} \right) \tilde{U}_{i \ell}^2
 \, = \, 
\frac{1}{n_1} \left( \WW^{\circ 2} \right)^t \, \tilde{\UU}_{i}^{\circ 2}\, .\\[.1in]
&\frac{1}{|A|} \sum_{j \in A} r_{ij} r_{iijj} 
\,=\, 
\frac{1}{n_1}\sum_{\ell = 1}^{n_1}  \left( \frac{1}{|A|} \sum_{j \in A} r_{ij} \tilde{U}_{j \ell}^2 \right) \tilde{U}_{i\ell}^2
 \, = \,
\frac{1}{n_1} \YY^t  \, \tilde{\UU}_{i}^{\circ 2}\, . \\[.1in]
&\frac{1}{|A|^2} \sum_{j,k \in A} r_{ij} r_{ik} r_{jjkk} 
\, = \,
\frac{1}{n_1}\sum_{\ell = 1}^{n_1}  \left( \frac{1}{|A|} \sum_{j \in A} r_{ij} \tilde{U}_{j \ell}^{\circ 2} \right)^2
 \, = \,
\frac{1}{n_1}\YY^t \, \YY \, . \\[.1in]
&\frac{1}{|A|^2} \sum_{j,k \in A} r_{ij} r_{ijjk}
\, = \,
\frac{1}{n_1}\sum_{\ell = 1}^{n_1}  \left( \frac{1}{|A|} \sum_{j \in A} r_{ij} \tilde{U}_{j \ell}^2 \right) \left( \frac{1}{|A|} \sum_{k \in A} \tilde{U}_{k \ell} \right) \tilde{U}_{i \ell} 
\, = \,
\frac{1}{n_1} \left( \WW \cdot \YY \right)^t \, \tilde{\UU}_i  \, .\\
\end{align*}

Thus, we can write $\hat{\tau}_1$ as a simple inner products of the vectors $\YY, \WW,$ and $\tilde{\UU}_i$,
\begin{align*}
n_1 \hat{\tau}_1
&\, = \,
\frac{1}{4} r_A^2 \one^t \tilde{\UU}_i^{\circ 4}
\, + \,
r_A \left[\frac{1}{2} \YY^t \, \tilde{\UU}_{i}^{\circ 2}
\, - \,
\WW^t \tilde{\UU}_{i}^{\circ 3} \right] 
\, + \, 
 \left( \WW^{\circ 2} \right)^t \,  \tilde{\UU}_{i}^{\circ 2}
\, - \,
\left( \WW \cdot \YY \right)^t \,  \tilde{\UU}_i 
\, + \,
\frac{1}{4} \one^t \, \YY^{\circ 2} \\[.1in]
\hat{\tau}_1 & \, = \,
\frac{1}{n_1} \sum_{\ell = 1}^{n_1}
\left\{ \frac{ r_A^2 }{4} \, \tilde{U}_{i \ell}^4
\, - \,
 r_A W_\ell \, \tilde{U}_{i \ell}^3
  \, + \,
\left( \frac{ r_A Y_\ell }{2} \, + \, W_\ell^2 \right) \tilde{U}_{i \ell}^2
 \, - \,
 W_\ell Y_\ell \, \tilde{U}_{i \ell}
  \, +  \,
 \frac{ Y_\ell^2 }{4}
\right\} \, .
\end{align*}
\qed \emph{ Lemma \ref{lem:varest} }

\begin{remark} \emph{We note that although the estimator $\hat{\tau}_1$ is consistent for a very general set of sampling distributions, it may in some cases converge slowly.  For very small sample sizes, we find the estimator to be negatively biased; that is, tests involving this estimator may be anticonservative.  Although the full DCM procedure appears in simulations to control false positive rate even for small sample sizes, we caution against its use when $\min(n_1, n_2) < 30$. } \end{remark}

\section{ Detailed Pseudocode }
\label{pseudocode}

\begin{algorithm}[H]
\setstretch{1.4}
\caption{ Initial Search Procedure }
\begin{algorithmic}[1]
\Procedure{InitDCM}{$\XX_1, \XX_2, k$}\Comment{Target output size $k$.}
\State $\FF_1, \FF_2 \leftarrow$ Fisher transformed correlation matrices of $\XX_1, \XX_2$.
\State $B \leftarrow$ index set of size $k$ chosen uniformly at random
\Repeat
\State $S =  \dsum_{i,j \in B} \left(\FF_1 - \FF_2\right)_{ij}$
\For{$a$ in $B$, $r$ in $B^C$}\Comment{Possible swaps}
\State $S_{ar} = \dsum_{\mathclap{i,j \in B\cup \{r\} \diagdown \{a\}}} \left(\FF_1 - \FF_2\right)_{ij}$
\EndFor
\State $a^*, r^* \leftarrow$  maximizers of $S_{ar}$ subject to $S_{a^*r^*} > S$.
\State $B \leftarrow B\cup \{r^*\} \diagdown \{a^*\}$ \Comment{Best swap}
\Until{no such $a^*, r^*$ exist}\\
\Return $B$
\EndProcedure
\end{algorithmic}
\end{algorithm}

\begin{algorithm}[H]
\setstretch{1.4}
\caption{Core Search Algorithm}
\begin{algorithmic}[1]
\Procedure{DCM}{$\XX_1, \XX_2, A$}\Comment{Initial index set $A$}
\State $A_{prev} \leftarrow \emptyset$
\State $cycle \leftarrow 0$
\Repeat
\State $t =$ 1 or 2 \Comment{Sample class label}
\State $m_t \leftarrow \text{mean}\left(\{\XX_{t}\}_A\right)$
\For{$i$ in $1, \cdots, p$}
\State $r_{ti} \leftarrow \widehat{\Cor}\left( {X_{ti}\, , m_t} \right)$
\State $\hat{T}_i \leftarrow m_1r_{1i} - m_2r_{2i}$ \Comment{Sample test statistic.}
\State $H_{0,i}:  T_i = 0$
\State $p_i \leftarrow P\left(T_i \geq \hat{T_i} \given H_{0,i}\right)$ \Comment{DC variable p-values}
\EndFor
\State $A_{next} = \{i \st H_{0,i} \text{ rejected by FDR controlled multiple testing}\}$
\If{ $A_{prev} = A_{next}$ }  \Comment{Check for cycles.}
\State $A_{prev} \leftarrow A$
\State $A \leftarrow A \cap A_{next}$
\State $cycle += 1$
\Else  \Comment{Update sets.}
\State $A_{prev} \leftarrow A$
\State $A \leftarrow A_{next}$
\EndIf
\Until{$A_{next} = A_{prev}$ or $A_{next} = \emptyset$ or $cycle = 2$} \\
\Return{A}
\EndProcedure
\end{algorithmic}
\end{algorithm}

\end{appendices}

\pagebreak
\bibliographystyle{apa}
\bibliography{MDC_Cites}

\end{document}